# Beyond the Higgs

Wan Ahmad Tajuddin Wan Abdullah

*High Energy Physics Research Group, Physics Department, Universiti Malaya, 50603 Kuala Lumpur, Malaysia*

**Abstract.** A Higgs-compatible boson has been observed at the LHC at CERN. We briefly review the role of the Higgs in particle physics and describe some of the current challenges in understanding the fundamental structure of the universe. Is there supersymmetry and is it instrumental in uniting gravity with the other three fundamental forces? What makes up dark matter and dark energy? We also report on the efforts in experimental particle physics by Malaysian collaborators to answer some of these questions.

## OBSERVATION OF A HIGGS-COMPATIBLE BOSON

A Higgs-compatible boson has been observed at the Large Hadron Collider (LHC) at CERN, Geneva, as was announced recently [1,2]. Protons colliding at TeV energies at LHC produces debris detected by the CMS and ATLAS detectors, among others. The CMS and ATLAS collaborators, each numbering almost 3000, from almost 200 institutions each from about 40 countries, sieve through about $10^{15}$ collision events to look for patterns in the product particles that signify decays of the boson.

According to standard theory, about 1 Higgs boson is expected in every $10^{10}$ events. The CMS and ATLAS collaborations did independent analysis, considering the possible decay channels of the Higgs, looking for the existence of the corresponding decay products, reconstructing the effective masses of the parents, and looking for any enhancements not accounted for by the established particle physics models nor by measurement noise.

Both experiments obtained enhancements at a mass of about 125 GeV/$c^2$. Channels explored are H → γγ, H → ZZ, H → WW, H → ττ and H → bb, with the first giving the most contribution, within errors. Channels with hadronic decay products are difficult to analyse due to the large hadronic background combinatorics. The decay products constrains the parent to be of integer spin, thus proving it at least to be a boson, if not the Higgs.

In this paper we review why this discovery is significant to the framework of particle physics, and what else is left for particle physics and cosmology.

## SIGNIFICANCE OF THE HIGGS

In the Standard Model (see e.g. [3]) of particle physics, three of the four fundamental forces, namely electromagnetism, weak interactions and strong interactions, are described as due to gauge boson exchange, basically between fermions which make up matter, by quantum field theories. For these theories to be renormalizable, the fermions and the intermediary gauge bosons should be massless. However, the gauge bosons responsible for the weak force, namely the W and the Z, are thought to be massive, and indeed measured as such, to explain finite range of the weak force.

To reconcile this, the W's and the Z are deemed to acquire masses through the Higgs mechanism [4-6]. This involves the concept of spontaneous symmetry breaking, when a system goes from a symmetric but metastable ground state in an e.g. quartic potential, to a stable one but sacrificing the symmetry. The Higgs field provides the potential, and the (local) symmetry breaking gives mass to the gauge bosons. The initially unified electroweak interaction [7] then breaks up [8,9] at low energies to the separate electromagnetic and weak interactions.

The Higgs particle, consequent to the Higgs field, has been searched for in the past decades (see e.g. [10] for a review), and thus the excitement when strong signs of its discovery was announced.

---

# CURRENT CHALLENGES

Would the discovery of the Higgs boson at CERN complete the Standard Model and close the chapter on particle physics, as well as on its implications in cosmology? Not in the very least!

The Standard Model is itself incomplete with regards to a theory to describe all the forces ('The Theory of Everything") since it strictly does not incorporate gravity, which is well described by the effect of space-time curvature (rather than by a quantum gauge theory), through General Relativity (see e.g. [11]). This description of gravity also allows the expansion of the universe, as in the Big Bang theory. This demands the interplay of particle physics, in the formation of structure and complexity as the universe cools [12]. Thus we have the Standard Model for cosmology, which can quantitatively predict things like the cosmic microwave background radiation and helium abdundance in the universe. Theories in particle physics then have their repercussions in cosmology, and vice-versa.

Even with the Standard Model for particle physics, not all have been understood. Firstly, even the nature of the Higgs itself has still a lot to be explored, e.g. the multiplet structure [13,14], suggested by the seemingly high diphoton decay rate, and its properties [15].

Also some questions remain, irrespective of the Higgs. While in principle the Standard Model is well-understood, some of its phenomenology are not calculable from first principles. The strong nuclear force, well-described perturbatively by Quantum Chromodynamics (QCD) (see e.g. [16]), for example, has not been able to quantitatively describe hadron structure, other than the state should be colourless (colour being the QCD charges). Baryons, with red, green and blue quarks in colour singlets, and mesons, quarks and antiquarks in colour-anticolour states are well-accepted, but could there be exotic states like tetraquarks, pentaquarks and hexaquarks? Also how coloured quarks 'fragment' to become colourless hadrons in jets are only given by models. How are baryons produced in fragmentation, for example.

In the weak sector, where weak states are described by 'flavours' of the fermions, things are more intriguing as it seems that the weak eigenstates do not coincide with the energy eigenstates. Therefore flavours can mix, as described by the Cabibbo-Kobayashi-Maskawa (CKM) mixing matrix [17,18] for the quark flavours. Neutrino flavours have also been discovered to mix (for a review, see [19]), so the question remains whether the charged leptons – electrons, muons, and tauons – also mix [20]. Flavour-changing neutrinos also mean that neutrinos need to have mass, which brings into question the description of the neutrino – whether Dirac, with distinct helicity states, or Majorana, with mixed left- and right-handed helicity states, giving rise to neutrinoless double beta decay. The CKM matrix also allows for charge-parity (CP) violation, an important ingredient for the cause of matter-antimatter asymmetry in the universe. Furthermore, the CPT theorem from quantum field theory predicts time-invariance violation if CP is violated, and this has indeed been seen [21].

Another unfinished business in the Standard Model is Grand Unification [22] – how the electroweak force is to be unified with the strong force. This would give rise to new gauge bosons, perhaps mixed lepton-quark states or leptoquarks, and also proton decay from the now-possible decay of d quarks to electrons. All these are yet to be seen experimentally.

A pressing theory just beyond that of the Standard Model for particle physics is supersymmetry, and experimental results testing this and the Standard Model for cosmology are that pertaining to dark matter and dark energy.

## Supersymmetry, Supergravity and Superstrings

Supersymmetry (for an introduction, see [23]) is the symmetry between fermions, particles with half-integer spins, and bosons, particles with integer spins. A motivation for supersymmetry is that local symmetry naturally leads to gravity ("supergravity") (see [24]). Gravitation is united to the other forces by describing it through quantum field theory as well. Another is that the running coupling constants of the electromagnetic, weak and strong interactions come together (unification) if supersymmetry is included [25], otherwise not. Supersymmetry also can answer the hierarchy problem, concerning why particles have such differing masses, by providing extra loop interactions with the new particles, the supersymmetric partners of existing ones.

Supersymmetry predicts the existence of supersymmetric partners ('superpartners') of existing particles, which are yet to be detected even at the highest energies at LHC (for a recent review, see [26]). Limits on their masses are getting more stringent, and the probability of their existence is getting smaller.

The converse way of uniting gravity with the other three forces is to describe this other forces as due to curvature of spacetime. Extra dimensions for these extra forces are required, and these extra dimensions are tightly-curled up

('compactified') so that effectively, spacetime is still 3+1 dimensions. Some theories predict detectably large extra dimensions, but these has not been seen either at LHC [27].

In superstring theories, supersymmetric particles in extra dimensions are extended to be string-like, bringing better mathematical behavior. At the moment, little of superstring theories are testable experimentally.

If the 125 GeV/c2 boson is a Higgs, some constraints apply to models extending the Standard Model with supersymmetry [28,29]. Minimal Supersymmetric Standard Model (MSSM), the simplest supersymmetric extension (with $N$=1 supersymmetry, i.e. only one supersymmetry operator) to the Standard Model requires fine tuning to fit the data, the nearly Minimal Supersymmetric Standard Model (nMSSM) is ruled out, while the Constrained Minimal Supersymmetric Standard Model (CMSSM) is disfavoured by the high diphoton rate measured. The Next-to-Minimal Supersymmetric Standard Model (NMSSM) fits the data naturally and actually gives a better fit than the Standard Model itself.

## Dark Matter and Dark Energy

Dark matter arises from the discrepancy between the amount of luminous mass in a spiral galaxy and its measured rotation rate (for a review, see [30]). Cosmology also suggests dark matter if the universe is to be flat. It can inferred that dark matter are non-baryonic (or else it would be luminous), and stable (no decays have been detected). The million-dollar question is what makes up this dark matter. Candidates include massive neutrinos and light superparticles.

Higgs bosons can be used to probe dark matter [31].

Measurements of a certain type of supernova allow one to measure astronomical distance more precisely and to notice that the expansion of the universe is accelerating. This is attributed to the existence of 'dark energy' (see [32] for a review), making up some 73% of the universe (dark matter 24%). Dark energy is still to be understood.

## MALAYSIAN EFFORTS

In Malaysia, we have started to work with the large international particle physics collaborations to contribute to these exciting physics, since about 2005.

With the ZEUS Collaboration (which ended data-taking in 2007), looking at electron-proton collisions at high energies on HERA at DESY, Germany, we worked on the electronics of the calorimeter read-out control [33], hadron calorimetry, fragmentation – baryons, strange correlations, pentaquarks and exclusive and diffractive particle production (exclusive ψ' photoproduction [34]).

For the high-energy frontier, we are working with the CMS Collaboration, which is running, on electromagnetic calorimetry, jets analysis, and grid computing. For the high-intensity frontier, looking for rare events, we work with the Belle II Collaboration on the SuperKEKB electron-positron collider in Japan, to start producing a trillion B meson pairs in 2014. Here we are interested in the drift chamber, muon detector, Monte Carlo simulations, cloud computing and rare exclusive events. We are also with the COMET Collaboration, just starting to form, to look for charged lepton flavor violation using high-intensity muon beams at JPARC, Japan.

## ACKNOWLEDGMENTS

Support by HIR and UMRG grants from Universiti Malaya, and an ERGS grant from Malaysian Ministry for Higher Education, is gratefully acknowledged.